\documentclass{svproc}
\usepackage{url}

\usepackage{graphicx}

\usepackage[colorlinks=true,linkcolor=blue,citecolor=blue,urlcolor=blue]{hyperref}

\usepackage{bbold}

\begin{document}
\mainmatter              
%

\title{Revealing correlations between a system and an inaccessible environment}

%
%

\author{Manuel Gessner\inst{1} \and Heinz-Peter Breuer\inst{2}}

%

\institute{D\'{e}partement de Physique, \'{E}cole Normale Sup\'{e}rieure, PSL Universit\'{e}, CNRS,
24~Rue Lhomond, 75005~Paris, France,\\
\email{manuel.gessner@ens.fr} 
\and
Physikalisches Institut, Universit\"at Freiburg,  Hermann-Herder-Stra{\ss}e 3, D-79104~Freiburg, Germany,\\
\email{breuer@physik.uni-freiburg.de}
}

\maketitle              

\begin{abstract}
How can we detect that our local, controllable quantum system is correlated with some other inaccessible environmental
system? The local detection method developed in recent years allows to realize a dynamical witness for correlations 
without requiring knowledge of or access to the environment that is correlated with the local accessible quantum system. 
Here, we provide a brief summary of the theoretical method and recent experimental studies with single photons and 
trapped ions coupled to increasingly complex environments. 
\end{abstract}

\section{Introduction}
Correlations are a ubiquitous concept in the field of quantum information theory. Establishing their presence is therefore a 
central task in many experimental studies and applications. Typically, this requires access to all of the correlated parties in 
order to perform a measurement of some observable which is sensitive to the correlations in question. In some 
cases, however, access may be limited to only one controllable quantum system that may share correlations with other 
parties, beyond the reach of the experimenter.

Most prominently, quantum correlations must be shared non-locally among several parties to realize, e.g., quantum 
communication protocols. Usually, experimental access for each party is limited to the local degrees of freedom whereas 
those of  other parties remain inaccessible. Similarly, interaction with an undesired eavesdropper can create correlations 
that may be harmful to the security of the protocol. Yet, the eavesdropper's system is not available for measurements. In realistic situations, quantum systems further become correlated with their environment due to the unavoidable 
interaction with uncontrollable modes. Also in this case, the environment is usually not accessible for measurements to 
verify the presence of these correlations. Finally, we may also consider the problem of characterizing a high-dimensional 
multipartite system. If it was possible to identify the correlation properties without requiring measurements on all of its 
subsystems, the complexity of the task could potentially be reduced dramatically. All of these scenarios lead us to the 
question, how can correlations with an inaccessible system be revealed?

In this manuscript, we will review the local detection method that permits to identify correlations in a bipartite system by only 
measuring one of the two subsystems. This is enabled by the strong dynamical influence of initial correlations on the 
evolution of an open quantum system \cite{Breuer2002}. More specifically, the detected correlations can be identified 
as quantum discord and the local signal can be used to provide a lower bound on a quantitative measure of this class of 
correlations.

The goal of the present contribution is to provide a brief overview of the recent theoretical and experimental activities on this 
topic. For a more thorough discussion of the technical aspects we refer to the original literature and the recently published 
review article \cite{GessnerChapter}, as well as selected Chapters in \cite{GessnerPHD}.

\section{Theory}

\subsection{General method}\label{sec:gen_meth}
A local witness for initial correlations between an accessible system and an inaccessible environment was first introduced in \cite{LaineEPL2010}. By monitoring the evolution of the trace 
distance \cite{Nielsen2000} between two arbitrary quantum states of the accessible system, a witness for initial correlations 
in either of the two initial states can be constructed. To this end, one makes use of the contractivity property of the trace 
distance \cite{RuskaiRMP1994} which ensures that the trace distance of any pair of states can never increase under 
positive maps, such as a dynamical evolution in absence of initial correlations. Hence, any increase of the trace distance 
above its initial value is an indicator of initial correlations (assuming that the environmental quantum state is the same in 
both cases) in any of the two quantum states \cite{LaineEPL2010}. This method allows for a direct information-theoretic 
interpretation and quantification of the information flow between the system and the environment and the correlations 
between them, in close relationship to measures for quantum non-Markovianity developed recently \cite{BreuerRMP2016}. 
However, it leaves open the question about how the two states should be chosen, e.g., in the case 
when the system is prepared in one given state and we are interested in its correlations. Furthermore, one may be 
interested in learning more about the specific class of correlated quantum states that is identified by a positive witness.

\begin{figure}[tb]
\centering
\includegraphics[width=.98\textwidth]{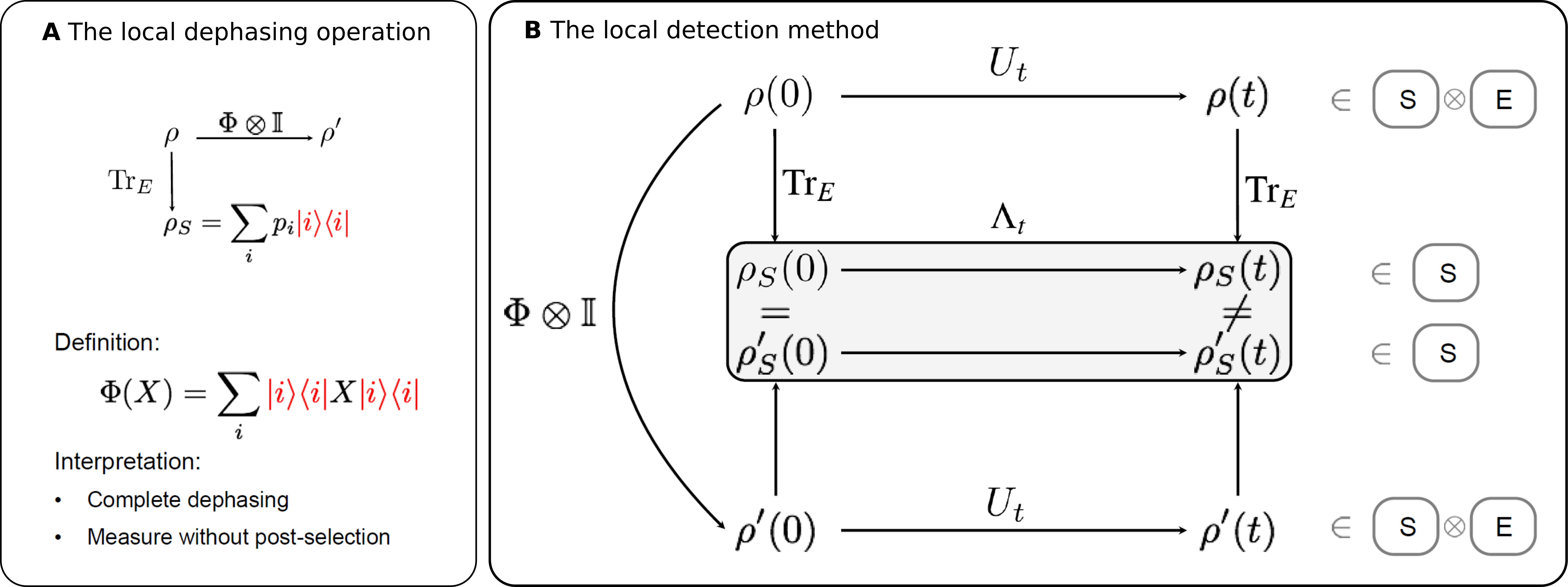}
\caption{The local detection method contains two central elements: 1. The application of a local dephasing 
operation $\Phi$ (panel A) and 2. the local dynamical evolution that depends on the correlations. Panel B summarizes the 
theoretical method \cite{GessnerChapter,GessnerPHD,GessnerNatPhys2014}. Subscript $S$ and $E$ refer to degrees of 
freedom of the accessible system and the inaccessible environment, respectively. Ignoring the environmental degrees of 
freedom is formally described by the partial trace operation $\mathrm{Tr}_E$. The combined dynamics of system and 
environment is described by the unitary map $U_t$ and the dynamical open-system evolution is represented by the map 
$\Lambda_t$.}
\label{fig:scheme}
\end{figure}

The local detection method extends the above witness for initial correlations to answer these questions, as has been demonstrated in
\cite{GessnerPRL2011}. The underlying scheme is represented in Fig.~\ref{fig:scheme}. We assume that the total system, 
which is only partially accessible, is initially in the quantum state $\rho$. The accessible degrees of freedom form the 
subsystem $S$ (regarded as open system), while the inaccessible degrees of freedom constitute the environment $E$. 
Furthermore, a quantum operation $\Phi$ acting locally on the accessible subsystem is used to produce a suitable 
reference state
\begin{equation}
 \rho' = (\Phi \otimes {\mathbb 1}) \rho.
\end{equation}
Taking $\Phi$ to be a controlled local dephasing operation on the accessible system (see Fig.~\ref{fig:scheme}), 
the reference state $\rho'$ becomes locally indistinguishable from the original state $\rho$, i.e., both states share the same 
reduced density matrices for system and environment. This means that if the two states $\rho$ and $\rho'$ 
are different, they must differ in terms of their correlations. Moreover, if the initial state $\rho$ contains no correlations it is 
easy to verify that the reference state $\rho'$ will be identical to $\rho$.

The potential change of correlations that is entailed by the local dephasing operation can have a significant impact on the 
dynamics of the accessible open system. Denoting the unitary time evolution operator of the total system 
(composed of open system $S$ and environment $E$) by $U_t$, we can write the reduced open system density matrices 
corresponding to the total initial states $\rho$ and $\rho'$ as follows,
\begin{equation}
 \rho_S(t) = {\rm{Tr}}_E\{U_t \rho U^{\dagger}_t\} \quad {\mbox{and}} \quad 
 \rho'_S(t) = {\rm{Tr}}_E\{U_t \rho' U^{\dagger}_t\}.
\end{equation}
While at time zero $\rho_S(0)$ and $\rho'_S(0)$ are identical, any deviation of the open-system states at some later
time, i.e. 
\begin{equation} \label{ineq-rhos}
 \rho_S(t) \neq \rho'_S(t) \quad {\mbox{for some}} \quad t>0, 
\end{equation} 
provides a witness for correlations in the initial state $\rho$. Notice that the local dephasing operation and the 
measurements on the local evolution can be realized without accessing the environment at any point. 

Introducing an appropriate norm in the open system's state space we can define a distance measure for
quantum states by means of
\begin{equation} \label{distance}
 d(t) = || \rho_S(t) - \rho'_S(t) ||,
\end{equation} 
such that condition (\ref{ineq-rhos}) can be written as $d(t) > 0$.
It is possible to interpret this scheme in the context of the witness discussed at the beginning of this chapter. In this case, 
we have chosen the pair of states as $\rho$ and $\rho'$, where the two are related to each other by the local dephasing 
operation. Since this operation never introduces correlations, this construction allows us to trace back any witness for 
correlations to the original state $\rho$. Furthermore, since both states have by construction the same initial reduced 
density matrix their initial distance is zero. A witness for initial correlations is thus registered when they become the 
least bit distinguishable. For this reason, the local detection method is not linked to a particular choice for a distance
measure for quantum states, such as the trace distance. Instead, it can be realized by any suitable observable that 
indicates the difference of the evolved quantum states at some later time $t$.

Let us finally also discuss the question regarding the nature of the detected correlations. It is possible to show that the 
distance between the states $\rho$ and $\rho'$ is a simple measure for discord-type correlations 
\cite{GessnerPRL2011,GessnerPRA2013,GessnerEPL2014}. 
Quantum discord describes a non-classical phenomenon that occurs in correlated bipartite quantum states  
\cite{ModiRMP2012,GessnerPRA2012}. Quantum states that do not commute with any local observable have non-zero 
discord \cite{GirolamiPRL2013}. While for pure states this concept is equivalent to entanglement, the two notions are 
different for mixed states. Quantum discord can be related to the performance of several quantum information protocols 
\cite{MadhokIJMP2013}, most notably the activation of entanglement \cite{StreltsovPRL2011,PianiPRL2011}, the 
distribution of entangled quantum states with a separable carrier 
\cite{CubittPRL2003,StreltsovPRL2012,ChuanPRL2012,FedrizziPRL2013,VollmerPRL2013,PeuntingerPRL2013}, and local 
quantum inferometry \cite{GirolamiPRL2014}.

\subsection{Performance of various distance measures}\label{sec:perf_dist_measure}

As discussed above the local detection method based on the dephasing map 
$\Phi$ works, at least in principle, for any choice of metric in the open system's state space. This is due to the fact that
the reduced initial states $\rho_S(0)$ and $\rho'_S(0)$ are identical and, hence, any metric is able to detect 
whether or not the open systems states will differ at some later time. However, different observables
or distance measures can have different sensitivities to reveal that $\rho_S(t)$ and $\rho'_S(t)$ indeed differ significantly 
from each other.

A natural metric on the quantum state space is given by the trace distance mentioned already, which is based on the
trace norm defined by
\begin{equation}
 || X || = {\mathrm{Tr}} \sqrt{X^{\dagger}X}.
\end{equation} 
The trace distance has the advantage that it is a contraction under trace preserving quantum operations, which
leads to the conclusion that the quantity defined in (\ref{distance}) provides a lower bound for the 
initial distance of the total states,
\begin{equation}
 d(t) \leq || \rho - \rho' ||,
\end{equation} 
and, hence, a lower bound for the discord-type quantum correlations mentioned above. 
Note that the right-hand side of this inequality is independent of time and, therefore, also the maximum 
over time represents a lower bound for such correlations:
\begin{equation}
 \max_{t\geq 0} d(t) \leq || \rho - \rho' ||.
\end{equation}

The trace distance between two quantum states $\rho_S$ and $\rho'_S$ can be interpreted as a measure for  the 
distinguishability of these states \cite{NielsenChuang,Hayashi}. This means that these states can be successfully distinguished by means of a single
measurement with a maximal probability given by
\begin{equation}
 p_{\max} = \frac{1}{2} \left( 1 + \frac{1}{2} || \rho_S - \rho'_S || \right),
\end{equation}
provided both states have been prepared with equal probabilities of $1/2$. 
In the case of a biased preparation of $\rho_S$ and $\rho'_S$ with different probabilities $p$ and $p'=1-p$, the
two states can be distinguished with a maximal probability of
\begin{equation}
 p_{\max} = \frac{1}{2} \left( 1 + || \Delta || \right),
\end{equation}
where $\Delta = p \rho_S - p' \rho'_S$ is known as Helstrom matrix \cite{Helstrom1976}. 

The performance of the local detection scheme based on the trace norm $||\Delta||$ of the Helstrom matrix has been 
studied in Ref.~\cite{Amato2018}. Quite interestingly, it turns out that for the method 
based on the dephasing map the trace distance, corresponding to the unbiased case $p=p'=1/2$, is optimal in the sense 
that it shows the largest increase due to the presence of correlations.

However, the situation changes if one considers detection schemes in which the reduced initial states are not equal
to each other. In fact, different distance measures can then show a quite different ability to detect 
initial correlations, as has been shown in Ref.~\cite{Wissmann2013}. On the one hand, the example studied in this reference indicates 
that the detection power of the trace distance is significantly larger than that of other well-known distance measures
for quantum states, namely the Bures metric, the Hellinger distance and the Jensen-Shannon divergence. 
On the other hand, in those cases the trace norm of the Helstrom matrix can show an even better performance 
than the trace distance \cite{Amato2018}.

\subsection{Applications to complex open quantum systems}
The method described above has been applied to the ensemble-averaged dynamics of complex open quantum 
systems in Refs.~\cite{GessnerPRL2011,GessnerPRA2013}, using group theoretical methods to determine
averages over the unitary group \cite{GessnerPRE2013} (see also \cite{CollinsCMP2006,ZnidaricPRL2011} for 
related techniques). 
Employing the Hilbert-Schmidt distance as a measure for the distance of quantum states,
it can be demonstrated that for a generic dynamical evolution, one expects the local detection method to successfully 
reveal correlations, but the influence of the initial correlations vanishes in the limit of systems with a large effective 
environmental dimension. Nevertheless, there are examples of memory-less, fully Markovian and infinite-dimensional 
environments that lead to the successful detection of correlations using the local detection method. For a more detailed 
discussion, see~\cite{GessnerChapter}. Further theoretical examples may also be found in~\cite{GessnerDA}.

\subsection{Application: Spin chain undergoing a quantum phase transition}
Quantum correlations play a special role for the ground state of quantum many-body systems undergoing a quantum phase 
transition \cite{AmicoRMP2008}. 
As a function of some external control parameter, the properties of the system change abruptly, most notably those of the 
ground state~\cite{SachdevBOOK2011}, but the transition usually affects the entire excitation 
spectrum \cite{CaprioAP2008,GessnerPRB2016}. 

Measuring these correlations is challenging due to the large number of degrees of freedom in extended many-body 
quantum systems. For this reason, it is interesting to notice that the local detection method allows us to reveal the drastic 
change of the ground-state correlation properties through measurements on only a single particle \cite{GessnerEPL2014}. 
In a theoretical study, a one-dimensional spin model with long-range interactions is used as a testbed for the local 
detection method. Measurements are performed only on one of the spins in the chain. The quantum phase transition is 
indicated by a peak in the signal related to the correlations between the measured spin and the bath formed by the 
remainder of the spins in the chain. It is remarkable that this signal is visible even for small, finite temperatures as 
demonstrated in Ref.~\cite{GessnerEPL2014}.

\section{Experiments}
\begin{figure}[tb]
\centering
\includegraphics[width=.98\textwidth]{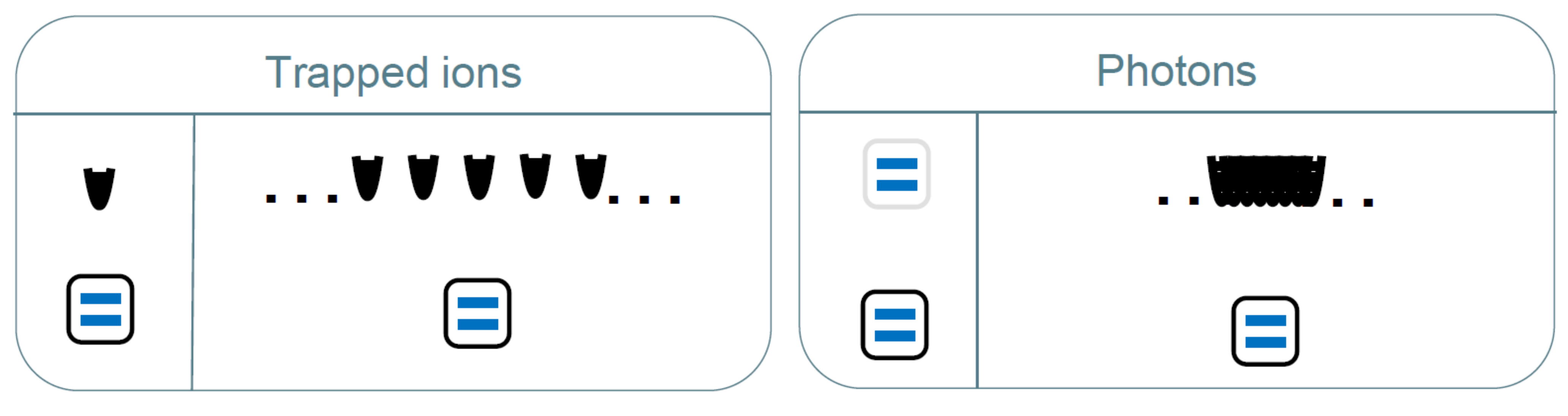}
\caption{Four classes of experimental scenarios in which the local detection method has been implemented. Two blue bars represent a qubit system and the parabolas reflect quantum mechanical harmonic oscillator modes. The black frame highlights the accessible system, which in all four cases was a qubit. With trapped ions, a qubit coupled to a single-mode oscillator was experimentally studied in \cite{GessnerNatPhys2014}. A qubit coupled to a string of up to 42 modes was studied in \cite{AbdelrahmanNatComm2017}. With photons, a qubit coupled to another qubit was considered in \cite{CialdiPRA2014}. A qubit coupled to a continuum of modes was analyzed in \cite{TangOPTICA2015}.}
\label{fig:scenarios}
\end{figure}
The local detection method has been implemented in different scenarios with both trapped ions and photons. The experiments can be classified according to the schema displayed in Fig.~\ref{fig:scenarios}. In all cases, the controllable (accessible) quantum system was modeled by a qubit (two-level system). The simulated environments range from single qubits to a continuum of harmonic-oscillator modes and a chain of 42 transverse phonons.

\subsection{Single trapped ion}
In a first experiment with a single trapped ion, an electronic two-level system realized the controllable open system \cite{GessnerNatPhys2014}. Interactions with the single-mode harmonic oscillator environment, the ion's motional degree of freedom, can be implemented by suitable laser control. By driving the qubit transition with a detuning equal to the trap frequency, excitations are coherently exchanged between the two degrees of freedom. This evolution leads to the generation of correlations between the qubit and the motion. Combined with controlled changes of the laser-cooling parameters, this provides access to a class of correlated probe states with a tunable environmental temperature.

To realize the local dephasing operation, a spectrally broad transition was addressed by far-detuned laser light, inducing a controllable ac-Stark shift that can be chosen such that any phase information between the qubit's ground- and excited states is deleted. Since the qubit is diagonal in the computational basis (as is confirmed by state tomography), this operation realizes the desired local dephasing. Subsequent driving of the sideband transition and monitoring of the qubit reveals the presence of correlations in the initial state. A lower bound for the initial quantum discord is obtained by evaluating the trace distance of the two evolutions, which can be directly extracted from the evolution of the excited-state population. The signal is visible also at higher temperatures, as is confirmed by experimental data at average phonon numbers up to around 5 \cite{GessnerNatPhys2014} and analytical arguments for even higher values \cite{GessnerPHD}.

\subsection{Chain of trapped ions}
\begin{figure}[tb]
\centering
\includegraphics[width=.98\textwidth]{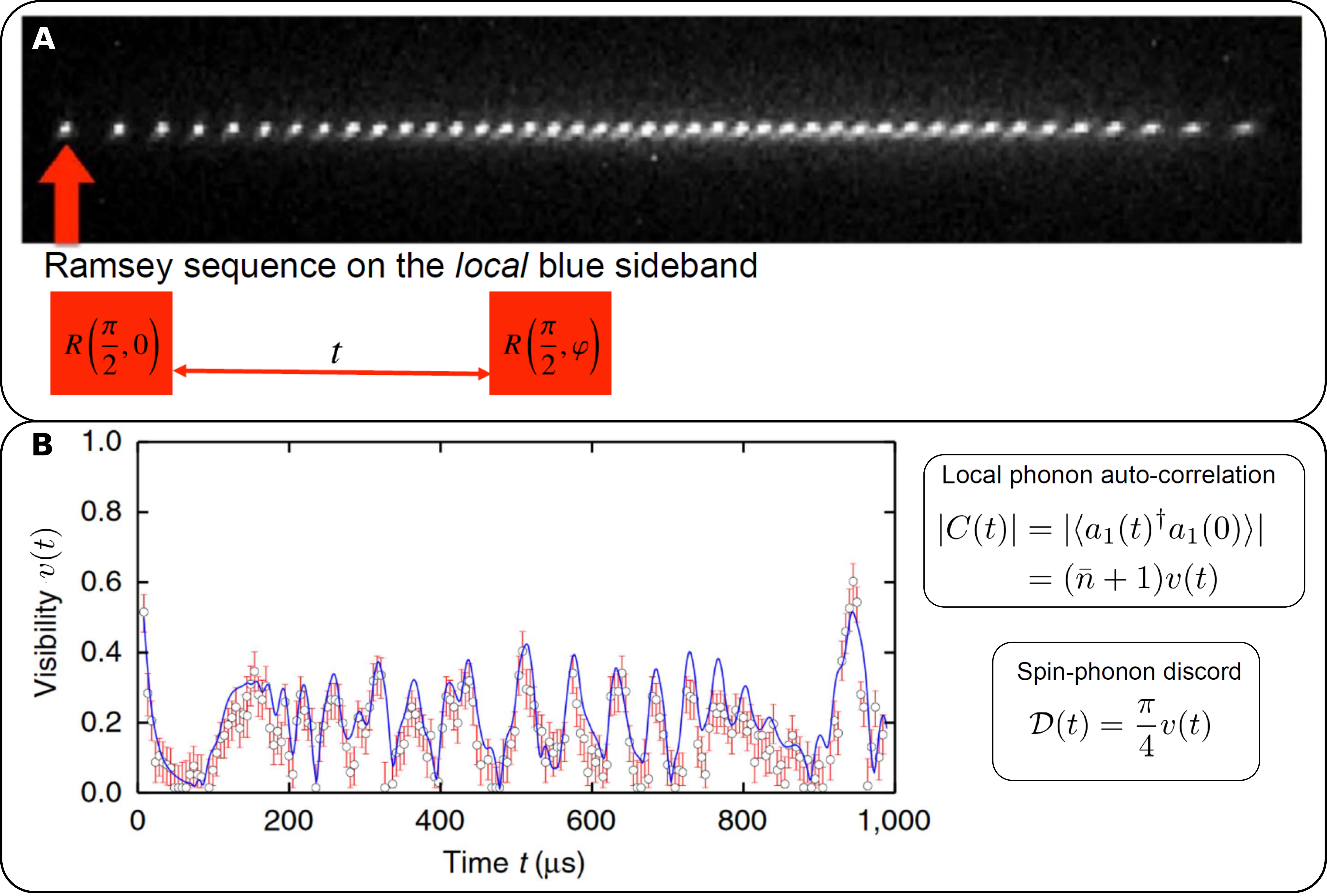}
\caption{A Ramsey sequence (A) is performed on the local sideband of the left-most ion in a chain of 42 ions \cite{AbdelrahmanNatComm2017}. The visibility of the Ramsey sequence (B) can be linked directly to the phonon auto-correlation function and the spin-phonon discord \cite{AbdelrahmanNatComm2017}.}
\label{fig:chain}
\end{figure}
In a later extension of the above experiment, a single electronic qubit is correlated with the phonon degrees of freedom of 
trapped-ion chains of variable length up to 42 ions \cite{AbdelrahmanNatComm2017}. All the phonons are coupled due to 
long-range Coulomb interactions. The experiment involves a Ramsey sequence on the sideband transition on a fast time 
scale, faster than the phonon-phonon hopping rate between neighboring ions (Fig.~\ref{fig:chain}). This ensures that the 
excitation of the qubit is accompanied by a creation of a localized phonon at the same site. As this local excitation is not an 
eigenstate of the chain, the phonon starts to travel and delocalize throughout the chain. By correlating the qubit with the 
single phonon that is created by the sideband interaction, the phonon can be traced during its evolution in the thermally 
excited bath of up to around 200 phonons, realizing a local quantum probe of a complex quantum dynamical system.

The revivals of the phonon at the initial site can be monitored through the visibility of a Ramsey interferometer sequence. 
To this end, two fast local sideband pulses are separated by a tuneable waiting time. Quite interestingly, it can be
shown within reasonable approximations that the locally measurable visibility $v(t)$ is directly linked to the modulus of the 
phonon auto-correlation function of the first site \cite{AbdelrahmanNatComm2017}
\begin{equation}
 \left| \langle a^{\dagger}_1(t) a_1(0) \rangle \right| = (\bar{n}+1) v(t).
\end{equation}
Moreover, the scheme allows the determination of a measure for the quantum discord ${\mathcal{D}}(t)$ between the 
electronic degree of freedom of the first ion and its motional degree of freedom, which is defined by the
change of the composite quantum state (measured in term of the trace distance) induced by the dephasing operation.
It turns out that within a very good approximation the discord is also directly related to the visibility through 
\begin{equation}
 {\mathcal{D}}(t) = \frac{\pi}{4} v(t),
\end{equation}
which enables the measurement of the quantum discord in the experiment \cite{AbdelrahmanNatComm2017}.

\subsection{Photons with a two-level environment}
Two environmental quantum states for photonic polarization qubits are realized by the paths at the output of a beam splitter 
\cite{CialdiPRA2014}. Using spontaneous parametric down-conversion and manipulating the polarization states as a 
function of their path leads to the controlled generation of polarization-momentum correlated photons. The discord-type 
correlations of these states can then be revealed using the local detection method without ever measuring the momentum 
degree of freedom. In this experiment, the method was complemented by a second step, a trace-distance based witness for 
initial correlations that is susceptible also to classical correlations of zero discord. This allows to identify and distinguish 
quantum discord from purely classical correlations.

\subsection{Photons with a continuum of environmental frequency modes}
In birefringent materials the polarization degree of freedom of single photons is coupled to the photon's own frequency 
modes \cite{TangOPTICA2015}. Each mode is described by a quantum harmonic oscillator and typical frequency 
distributions of single photons comprise a continuum of modes. This effect was harnessed to simulate a continuous, 
memory-less environment for optical polarization qubits for an implementation of the local detection method 
\cite{TangOPTICA2015}. 

A series of correlated states is prepared by Alice and sent to Bob whose task is to detect the presence of correlations 
without ever measuring the frequency modes. A Michelson-Moreley interferometer is used to reveal these correlations. The 
local dephasing operation is realized by means of a long polarization-maintaining fiber that destroys the phase information 
relative to polarization and frequency. Its axis is aligned to the measured eigenbasis of the polarization qubits to ensure 
dephasing in the correct basis. State tomography of the polarization qubit is performed before and after the dephasing. The 
correlations are successfully revealed using the local detection method, even though the coupling of the open system is 
realized with a continuum of modes that represent a fully Markovian environment.

\section{Conclusions}
The local detection scheme discussed in this contribution provides a method to locally detect and quantify correlations 
between an accessible open system and an inaccessible environment. Thus, in more general terms it allows to determine 
correlations in a composite quantum system without requiring access to all of its subsystems. 
Necessary requirements for implementing the method are (i) the presence of interactions between the potentially 
correlated subsystems, and (ii) a good level of control of the accessible part of the composite system. 
The second condition refers to the implementation of the local dephasing operation, which requires knowledge 
of the eigenbasis of the state of the accessible subsystem.

As we have discussed and illustrated here, the scheme is very general and flexible, and paves the way for many theoretical 
and experimental applications in the fields of complex open system and quantum information. In particular, it is
important to note that the scheme does not require control or even knowledge of the state of the total system, of the 
system-environment interaction Hamiltonian, or of the initial environmental state.

Up to now, experimental realizations of the local detection method have been carried out for both trapped ion systems 
and for photonic systems. In all these experiments the accessible, open system represents a qubit, formed by
an electronic degree of freedom of an ion or by the polarization degree of freedom of a photon. On the other hand,
the environmental system can either be another simple qubit system or a much more complex system formed by the
many modes of a long ion chain or by a continuum of frequency modes. 

The development of experimental applications to composite local systems would be highly interesting. It would allow to study the impact of correlations with an external environment onto entangled states of well-controlled degrees of freedom. This would further open up new avenues towards a theoretical extension of the local detection method to multipartite scenarios. A related recently developed method allows to detect quantum discord with an inaccessible system by witnessing the generation of entanglement among two non-interacting, controllable systems \cite{K1}.

Revealing nonclassical properties and correlations with inaccessible objects may provide a promising route towards identifying quantum effects in complex situations where a detailed quantum description of the object is challenging. This approach has been recently suggested in the context of biological systems \cite{K2}, quantum processes \cite{K3}, and even for tests of quantum gravity \cite{K4}.

\section*{Acknowledgments}
This work was funded by the LabEx ENS-ICFP:ANR-10-LABX-0010/ANR-10-IDEX-0001-02 PSL*. 
M. Gessner would like to thank the organizers of the 684. WE-Heraeus-Seminar ``Advances in open systems and fundamental tests of quantum mechanics'' for being given the opportunity to present this work.

\end{document}